# Dissociation of hydrofluorocarbon molecules after electron impact in plasma.

*Dmitry V. Makhov[1,2], Gregory Armstrong[3], Hsiao-Han Chuang[1,2], Harin Ambalampitiya[3], Kateryna Lemishko[3], Sebastian Mohr[3], Anna Nelson[3], Jonathan Tennyson[3,4], and Dmitrii Shalashilin[1]*

[1]*School of Chemistry, University of Leeds, Leeds LS2 9JT, United Kingdom*
[2]*School of Mathematics University of Bristol,* Fry Building, Woodland Road, Bristol, BS8 1UG, United Kingdom
[3]*Quantemol Ltd., 320 City Road, The Angel, London EC1V 2NZ, United Kingdom*
[4]*Department of Physics and Astronomy, University College London, London WC1E 6BT, United Kingdom*

***Abstract***

The process of dissociation for two hydrofluorocarbon molecules in low triplet states excited by electron impact in plasma is investigated by *ab initio* Molecular Dynamics (AIMD). The interest in dissociation of hydrofluorocarbons in plasma is motivated by their role in plasma etching in microelectronic technologies. Dissociation of triplet states is very fast, and the reaction products can be predicted. In this work, it was found that higher triplet states relax into the lowest triplet state within a few femtoseconds due to nonadiabatic dynamics, so that the simplest *ab initio* MD on the lowest triplet state seems to give a reasonable estimate of the reaction channels branching ratios. We provide evidence for the existence of simple rules for the dissociation of hydrofluorocarbon molecules in triplet states. For molecules with a double bond, the bonds adjacent to it dissociate faster than the other bonds.

## *1.* Introduction

Motivated by the search for new environmentally friendly molecules for use in plasma technologies, we study the process of dissociation of hydrofluorocarbon molecules in low energy triplet states after the electron impact in plasma. Hydrofluorocarbons and other fluoroorganic molecules are used in the microelectronics industry to generate free radicals for plasma etching, one of the main technological processes in microchip production. Many of these molecules and the products of their chemistry in plasma are damaging for the environment. They have high Global Warming Potential (GWP) and can also cause unwanted reactions in the Earth's atmosphere.

Breakup patterns of molecules as a result of photodissociation has been well studied theoretically. See for example reviews [1-3], where references to numerous applications can be found. Similarly, a number of recent studies have focused on developing theories of break-up patterns follow electron impact ionization [4-7]. This work is aided by extensive data from mass spectroscopy. Fragmentation patterns following electron impact dissociation are much less well studied. Ziolkowski *et al* [8] used a trajectory hopping approach to study the fragmentation patterns of methane following an R-matrix calculation of electron impact excitation but there appears to no work on the fragmentation of larger systems.

Here we investigate the dissociation of two hydrofluorocarbons, 1,1,1,3,3,3-Hexafluoropropane ($C_3H_2F_6$) and 1,3,3,3-Tetrafluoropropene ($C_3H_2F_4$) by *ab initio* molecular dynamics simulations using *Ab Initio* Multiple Cloning (AIMC) method [9, 10], in which an ensemble of Ehrenfest trajectories describes the nuclear motion of the electronically excited molecule. AIMC is somewhat more rigorous way to treat nonadiabatic transitions than surface hopping employed in [8]. It has been shown that AIMC can provide accurate description of nonadiabatic dynamics from first principles [10-15]. The Quantemol Electron Collisions (QEC) code [16], that interfaces with the UKRmol+ suite of molecular R-matrix codes [17] is used to determine the initial triplet states created by electron impact in plasma. Although AIMC was originally developed for the simulation of non-adiabatic dynamics of excited molecules in singlet states in photochemistry, it can equally be used for the dynamics of molecules in low energy triplet states produced by electron impact. Our study concentrates on these triplet states because they are lowest electronic states often separated from other excited states by a substantial energy gap.

Our simulations yield the dissociation kinetics of the molecular bonds along with branching ratios for various dissociation channels, which produce neutral free radicals. The branching ratios are important for understanding chemical composition of plasma, but very difficult to measure experimentally. The calculations appear to yield very simple rules that can be used to predict dissociation channels even without calculations. Our results show, in particular, that single bonds adjacent to the double bond in $C_3H_2F_4$ break more efficiently than other bonds.

## 2. Theory

### 2.1 Excitation via electron impact in plasma and subsequent dissociation channels.

In the recent decade, many quantum and classical molecular dynamics methods have been developed to simulate dynamics of molecules in excited electronic states, when electronic excitation energy is transferred into the energy of nuclei and often results in dissociation. The methods have been implemented in a number of codes [18-20] and have been applied to a number of photochemical reactions, where the absorption of a V-UV photon excites the molecule into a singlet state [1]. The interest in the dynamics of singlet states have been justified by their importance in many processes in photochemistry and photobiology. However, the nature of the initial excitation is not really important, and the same technique can be applied to the molecules excited by electron impact as well, where triplet electronic states are also produced. Indeed, as triplets are usually the lowest excited states, their dynamics are the most important in plasma. Following the initial excitation, detailed information about various channels of dissociation after electron impact can be obtained and used in the modelling of chemistry in plasmas. If the cross section $\sigma_{E_i}$ of excitation by electron impact of the molecule into a particular electronic state $E_i$ is known, then the cross section of dissociation into a channel $a$ can be easily calculated as

$$\sigma_{E_i a} = \sigma_{E_i} P_{E_i a} \qquad (1)$$

where $P_{E_i a}$ is the probability of the chemical channel $a$ for the molecule excited into electronic state $E_i$, which is calculated by molecular dynamics. In its simplest form $P_{E_i a}$, also called branching ratio, is a fraction of molecular trajectories ending in the appropriate dissociation channel $a$.

Figure 1 shows the cross section of excitation via electron impact for $C_3H_2F_6$ and $C_3H_2F_4$ molecules calculated with R-matrix method implemented in Quantemol software [16]. The ground-state configuration of $C_3H_2F_4$ is $(1-20a', 1-8a'')^2$. The excitation cross sections were calculated at the CASSCF(6e, 6o)/cc-pVDZ level of theory, using 19-22a' and 8-9a'' orbitals in the active space. An R-matrix sphere of radius 10 Bohr was used. For $C_3H_2F_6$, the ground state is $(1-13a_1, 1-7b_1, 1-11b_2, 1-6a_2)^2$. The cross sections were calculated at the CASSCF(8e,8o)/6-311G** level of theory, with an active space consisting of the 13-15$a_1$, 6-8$b_1$ and 11-12$b_2$ orbitals. An R-matrix sphere radius of 10 Bohr was found to be sufficient. It can be seen that, while the excitation threshold for the lowest triplet state in $C_3H_2F_4$ is separated from that for the rest of excited states by a large energy gap of about 5 eV, two lowest triplet states in $C_3H_2F_6$ molecule have similar excitation thresholds, and even more triplet states can be excited by electrons with the energies just 2eV higher. Therefore, in principle, many electronic states should be taken into consideration.

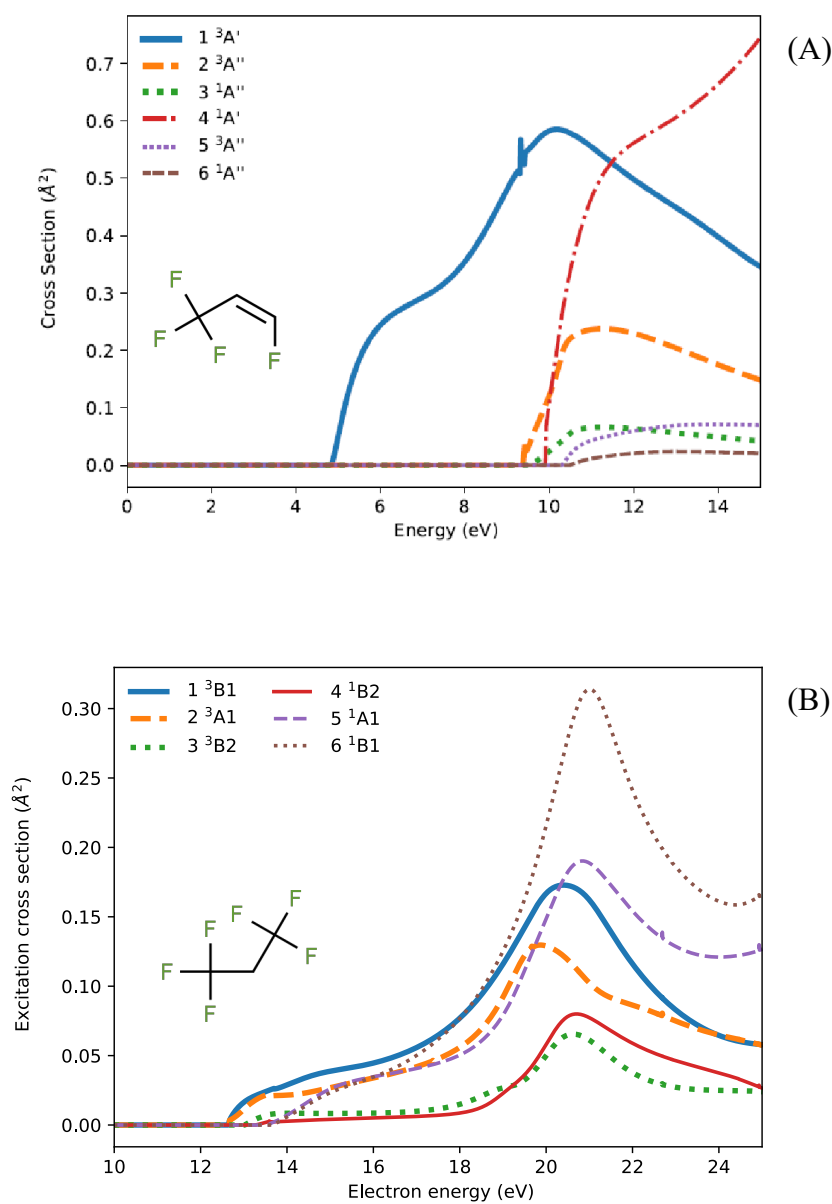

**Fig 1.** Electron impact excitation cross sections. (A) 1,3,3,3-Tetrafluoropropene, where low triplet state dominates at the electron energies of interest; (B) 1,1,1,3,3,3-Hexafluoropropane, where multiple states contribute for lowest energies.

## 2.2 Ab Initio Multiple Cloning – a method for simulating excited state non-adiabatic molecular dynamics.

In our *Ab Initio* Multiple Cloning (AIMC) approach, we run direct quantum non-adiabatic molecular dynamics simulations. The dynamics includes several Born-Oppenheimer electronic states with population transfer between them due to non-adiabatic coupling. In the direct dynamics approach, the trajectories of nuclear motion are run without the need for a precalculated potential energy surfaces. Instead, the potential energies of the Born-Oppenheimer electronic states, their gradients, and couplings between them are calculated at every time-step an using electronic structure package (Q-Chem in this case). An ensemble of such trajectories can simulate efficiently the dynamics of electronically excited molecule.

Our approach is based on Ehrenfest trajectories, which are guided by state-averaged force

$$\mathbf{F} = -\sum_I |a_I|^2 \nabla V_I + \sum_{I \neq J} a_I^* a_J \mathbf{C}_{IJ}(V_I - V_J), \quad (2)$$

where $a_I$ are quantum amplitudes for electronic states, $V_I$ are potential energies, and $\mathbf{C}_{IJ}$ is a non-adiabatic coupling vector. The second term here is the so-called Hellmann-Feynman force, which is associated with non-adiabatic electronic population transfer. The quantum amplitudes of electronic states are propagated together with coordinates and momenta of nuclei as:

$$\dot{a}_I = -\frac{i}{\hbar} \sum_J H_{IJ} a_J \quad (3)$$

where $H_{IJ}$ is the electronic Hamiltonian:

$$H_{IJ} = \begin{cases} V_I & I = J \\ -i\hbar \, \dot{\mathbf{R}} \cdot \mathbf{C}_{IJ} & I \neq J. \end{cases} \quad (4)$$

In principle, our AIMC approach can be made fully quantum, as the trajectories here are used just to guide nuclear Gaussian basis functions, and then the time-dependent Schrödinger equation is solved in the basis of trajectory guided nuclear basis functions (see our reviews [10, 21]). We have developed a number of sampling techniques that can be used to describe the initial conditions of the dynamics and the branching of trajectories in the region of nonadiabatic coupling where the transitions between electronic states occur. The latter is particularly important because Ehrenfest trajectories can become unphysical when several uncoupled electronic states with significantly different gradients have considerable populations. To address this issue, we apply a cloning procedure that results in the branching of trajectories, which reflects the bifurcation of wave-function at conical intersections (see [10, 21] for the details). Each branch gets a weight according to the appropriate Ehrenfest amplitudes. With a sufficient number of coupled basis functions guided by Ehrenfest trajectories, our approach can be converged to the exact fully quantum result.

For simulation of the dynamics after excitation via electron impact, however, we probably will not need the fully quantum approach with quantum coupling between the trajectories. We will run a number of independent uncoupled trajectories with random initial coordinates and momenta starting from all possible electronic states. Then, we identify the dissociation products and get the statistics for various dissociation channels. The probability $P_{E_i a}$ will then be simply given as the fraction of trajectories which end up in the channel $a$ weighted by their branching probabilities, if necessary. If only one electronic state is involved, AIMC becomes equivalent to standard *ab initio* MD in this case.

## *2.2 Electronic structure theory used.*

The choice of electronic structure theory method is crucial for the direct dynamics. Previously, in our simulations of photochemistry, we used CASSCF approach, which unfortunately proved to be unstable for direct dynamics on molecules of interest in plasmas, where the molecules experience strong deformations in the course of dynamics due to high temperatures.

In the direct dynamics approach, the electronic structure is called at every time step of the nuclear dynamics simulation, so the selected electronic structure methods must be cheap, and, at the same time, be able to capture most of the dynamic and static correlations during the propagation. After trying several electronic structure methods, we adopted Spin-Flip method TDDFT [22] implemented in the Q-Chem package [23] for our AIMC direct dynamics. The electronic structure theory used in dynamics was different from that used in the Electronic scattering R-matrix calculations. However, as only a few lowest electronic states are considered, we believe that their nature is the same in both approaches.

It is well known [24] that conventional linear-response TDDFT can only deal with single-electron excitation and cannot deal with dark states, multi-excitations etc. Furthermore, it cannot describe sizable hole/electron spatial separation and, also, cannot deal with degeneracies or near-degeneracies situations such as dissociations, diradicals, transition states and conical intersections. Spin-flip TDDFT method covers [25] most of these TDDFT difficulties by using high-spin state reference state instead the most stable Kohn-Sham ground state. The spin-flip ansatz applies a spin flipping excitation operator on a single high-spin reference state generating proper configurations for describing the excited states. This yields accurate potential energy surfaces of the excited states for the direct molecular dynamics.

**3.     Computational details and results.**

Our first aim is to estimate the importance of non-adiabatic effects in the process of the triplet state dissociation. We have chosen $C_3H_2F_6$ molecule as an example because its two lowest triplet states are very close to each other energetically, and have similar excitation cross sections. We run non-adiabatic dynamics for the two lowest triplet states using our AIMC approach with energies and forces provided by Q-Chem [23] at Spin-Flip TDDFT level using 6-31+G* basis set and BHHLYP hybrid functional.

The initial conditions are generated for T=5000 K using classical Boltzmann distribution for all vibrational modes, with lower temperatures investigated afterwards. As the atomic displacements at high temperature would be too large, beyond the applicability of harmonic approximation, we put all thermal energy into the kinetic energy when generating random initial momenta, and use the equilibrium geometry as an initial geometry for all trajectories. This should not cause any problems, as the energy redistribution between kinetic and potential energies happens very fast, within a few vibrational periods. We run 200 initial trajectories starting from both the lowest and second lowest triplet states with the number of branches growing in the course of the dynamics as a result of cloning.

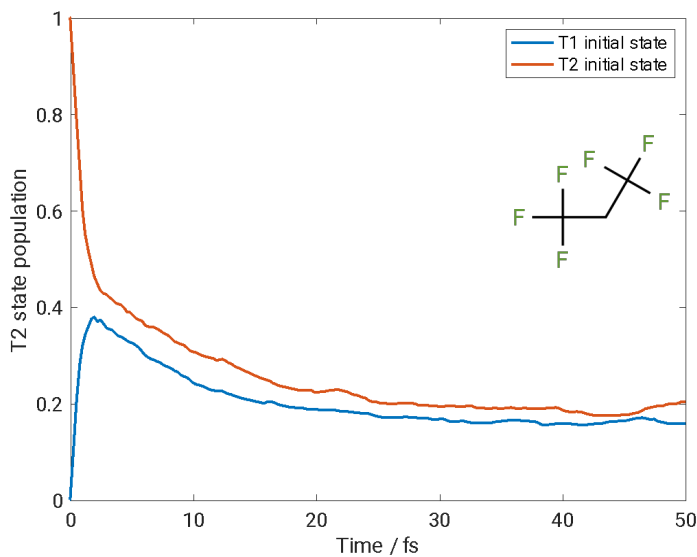

**Fig.2** The average populations of the upper triplet state of $C_3H_2F_6$ molecule as a function of time given by AIMC non-adiabatic dynamics calculations that use lower (blue) and upper (red) triplet states as an initial state.

Figure 2 shows the average populations of electronic states as a function of time. One can see very fast non-adiabatic population transfer in the very beginning of the dynamics: regardless of the initial state, the population of the lower triplet state became around 75% within just few femtoseconds. This suggests that we can avoid very expensive non-adiabatic calculations and run the dynamics only for the lowest triplet state. In order to test this suggestion, we run the dynamics for the molecule in the lowest triplet state without non-adiabatic effects taken into account and compare the results with those produced by AIMC calculations. One can see from Fig. 3 that there is little difference between the dissociation kinetics produced by this simple molecular dynamics on a single lowest triplet state potential energy surface and the AIMC dynamics, even when the latter starts from the upper triplet state. Some difference is found only for C-H bond dissociation where AIMC dynamics gives slightly higher dissociation yield; this can be explained by extremely fast C-H bond dissociation, which happens even faster than the molecule relaxation to the lower state.

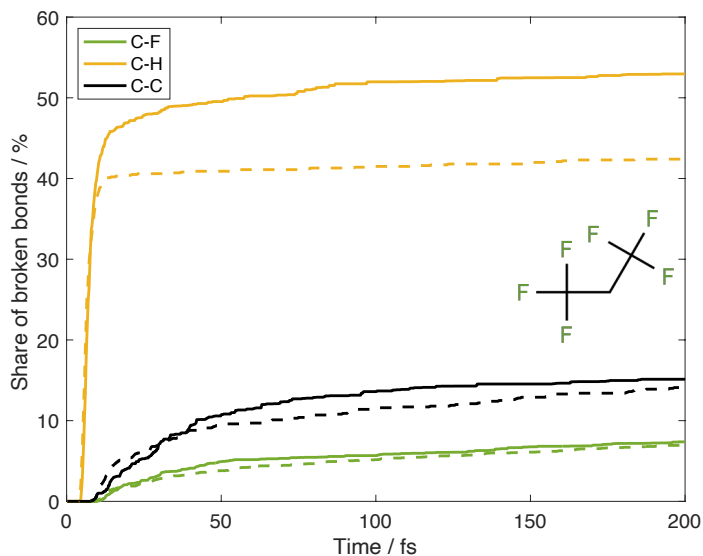

**Fig.3** The comparison of dissociation kinetics of $C_3H_2F_6$ molecule in a triplet state given by AIMC non-adiabatic dynamics simulations with initial state T2 (solid) and by the molecular dynamics on T1 potential energy surface (dashed). The number of broken bonds as a function of time.

Thus, the simulation of the dissociation in plasma in many cases does not require the use of the expensive non-adiabatic dynamics, and the essential details of the process can be captured by running the dynamics on the lowest triplet potential energy surface (either because only one state is initially excited, as in the case of $C_3H_2F_4$, or because the molecule quickly relaxes to the lowest triplet state, as in $C_3H_2F_6$). The rapid relaxation of triplet excitation was also found for methane [8], where the same conclusion about importance of lowest triplet state was made when computing fragmentation patterns. Surely, more calculations for other molecules are needed before any general conclusions can be made.

Considering the above, we run the extensive calculations for $C_3H_2F_6$ dissociation using cheaper single potential energy surface dynamics that include only the lowest triplet state. We study the dissociation kinetics at T=5000 K, 3000 K, 2000 K and 1000 K; for each temperature, we run an ensemble of 500 trajectories for 600 fs. The bond is considered broken when its length exceeds the 5 Å threshold.

Figure 4 shows the dissociation kinetics for all types of bonds at different temperatures. About half of the C-H bonds break very fast suggesting that electronic excitation is located at these bonds. The dissociation yield for C-H bond practically does not depend on temperature, which is consistent with the barrierless dissociation. Then, the gradual excitation transfer leads to the breaking of C-C bonds, followed by slow C-F bonds breaking.

The character of C-C bond breaking kinetics is somewhat in between that for C-H and C-F bonds. About 40 % of C-H bonds breaks within first 5 fs and then the dissociation essentially stops. For C-C bonds, about 5 % of them break fast but later than C-H bonds; then the dissociation gradually continues, at least at higher temperatures. Unlike the C-H bonds, the dissociation of C-F bonds mostly happens gradually, and only a minor fraction of them breaks fast within first 10-20 fs of the dynamics.

Table I lists the identified fragments for $C_3H_2F_6$ molecule at the end of the dynamics at different temperatures. The results are given for 500 molecules. One can see that there are no undissociated molecules, even at 1000 K. The dominating small radicals are H, F, $CF_2$ and $CF_3$. The large number of $C_3HF_6$ radicals, which is nearly equal to the number of H radicals at 1000 K and 2000 K, indicates that the dissociation of a C-H bond essentially eliminates the probability of any other dissociation at lower temperatures. The large number of $CF_2$, $CF_3$ and $C_2H_2F_3$ radicals at all temperatures shows that a significant number of F radicals are produced by the dissociation of C-F bond in $CF_3$ radicals that were produced by earlier C-C bond dissociation.

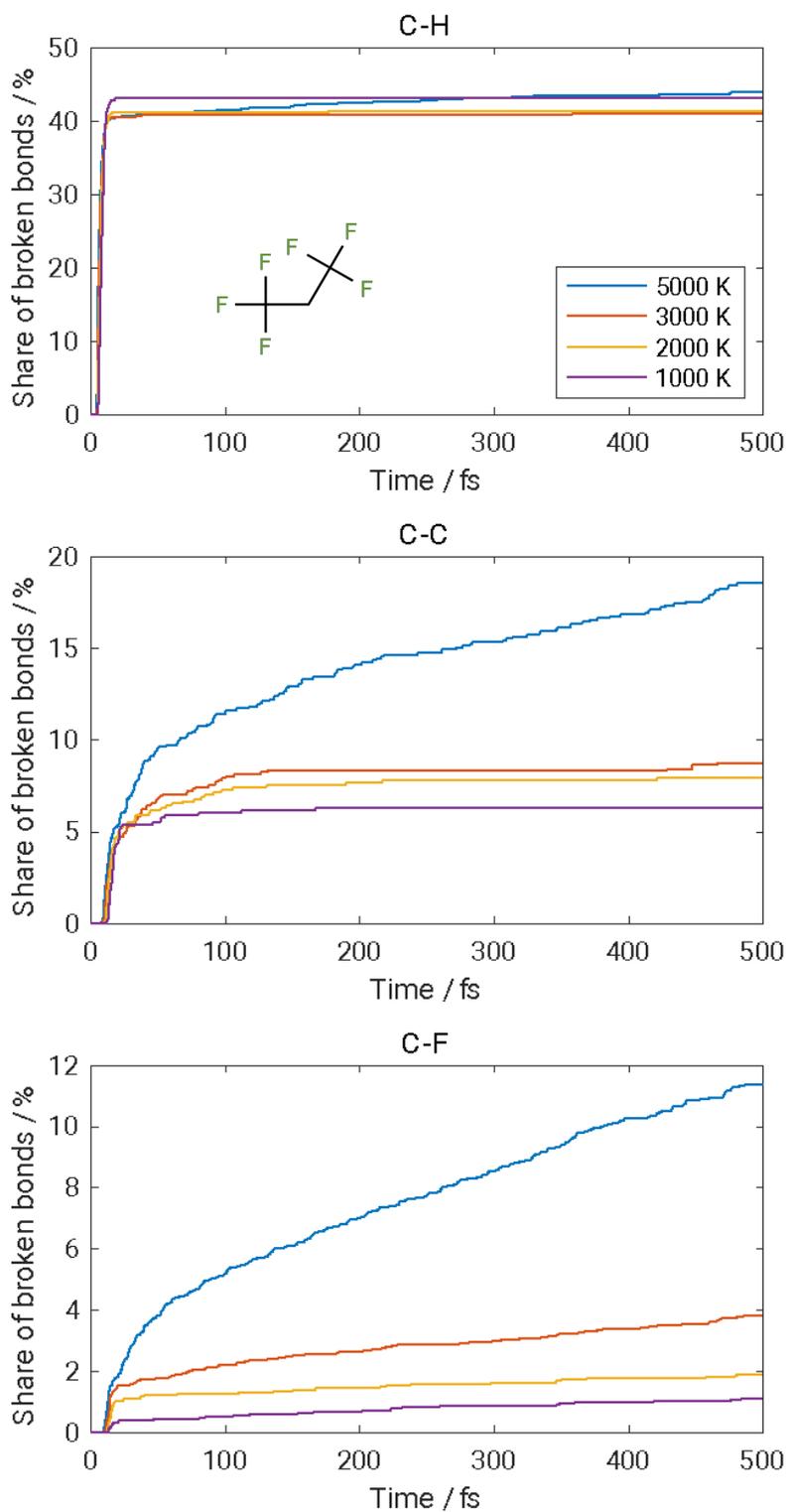

**Fig.4** The number of C-H (upper), C-C (middle) and C-F (lower) bonds broken as a function of time for different temperatures for $C_3H_2F_6$ molecule.

|        | **5000 K** | **3000 K** | **2000 K** | **1000 K** |
|--------|------------|------------|------------|------------|
| H      | 410        | 407        | 407        | 424        |
| F      | 308        | 125        | 70         | 40         |
| C3HF6  | 195        | 370        | 400        | 420        |
| C3HF5  | 112        | 31         | 6          | 1          |
| CF2    | 104        | 51         | 41         | 26         |
| CF3    | 62         | 33         | 42         | 43         |
| C2H2F3 | 33         | 56         | 61         | 63         |
| CF     | 29         | 7          | 1          | 1          |
| HF     | 28         | 2          | 3          |            |
| C2HF3  | 28         | 5          | 2          | 1          |
| C2H2F2 | 17         | 17         | 6          | 2          |
| CH2    | 15         | 4          | 6          | 1          |
| C3F5   | 15         |            |            |            |
| C2HF2  | 14         | 2          | 1          |            |
| CHF    | 11         |            |            |            |
| CH2F   | 11         | 1          | 1          | 1          |
| C2HF4  | 7          | 1          |            |            |
| C3H2F5 | 7          | 12         | 16         | 10         |
| C3HF4  | 6          |            | 1          |            |
| C3F4   | 5          |            |            |            |
| C2HF   | 4          |            |            |            |
| C3H2F4 | 4          | 1          |            |            |
| H2     | 3          |            |            |            |
| C2F2   | 3          |            |            |            |
| C2F3   | 3          |            |            |            |
| C3F6   | 3          |            |            | 1          |
| C      | 1          |            |            |            |
| CF4    | 1          |            |            |            |
| CH     | 1          |            |            |            |
| C2F4   | 1          |            |            |            |
| C2H2   | 1          |            |            |            |
| C2H2F  | 1          |            |            |            |

**Table I.** The number of fragments for $C_3H_2F_6$ dissociation per 500 molecules after 600 fs of the dynamics at different temperatures.

Figure 5 shows the dissociation kinetics for $C_3H_2F_4$ at 5000 K. The bond that breaks most efficiently here is the single C-C bond: about 40% of them are broken within 600 fs.

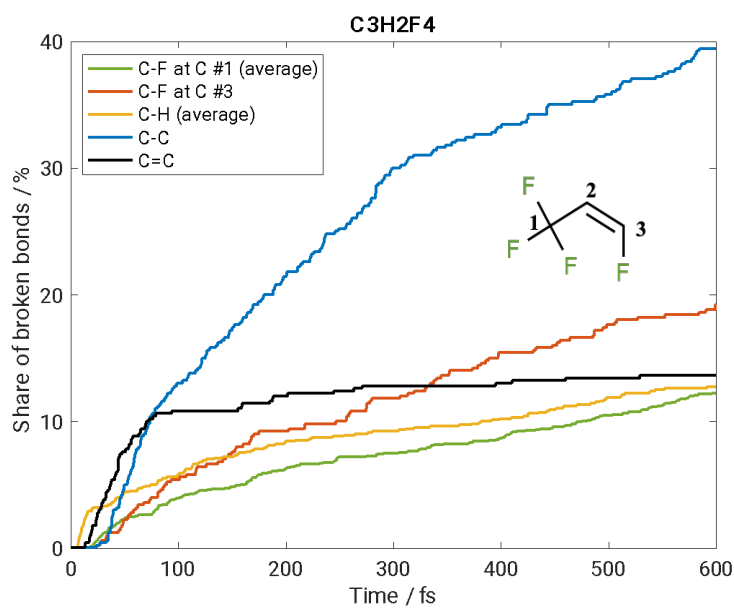

**Fig. 5.** The dissociation kinetics for $C_3H_2F_4$ molecule at 5000 K.

The character of C-H bond breaking kinetics in $C_3H_2F_4$ molecule is very different from $C_3H_2F_6$ case: C-H bonds here break gradually with approximately the same rate as C-F bonds. On the other hand, the kinetics of double C=C bonds breaking resembles that for C-H bond in $C_3H_2F_6$ molecule: the bonds break very intensively in the first 30 femtoseconds of the dynamics, then the process essentially stops. This agrees with the suggestion that the electronic excitation is initially localized at the double bond: the lowest triplet state results from the removing an electron from bonding π orbital and adding an electron to π* antibonding orbital. This state is isolated from all other excited states by a 5 eV energy gap. Therefore, *ab initio* MD on this single state are well justified even without additional simulations of non-adiabatic dynamics.

Comparing the dissociation rate for different C-F bonds, we have found that the bond at C atom involved in double bonding breaks more efficiently than C-F bonds at the other C atom. This can be explained by the initial localization of the excitation at double C=C bond, which then gradually spreads to the neighbouring bonds. This observation is extremely important, as it can help to establish simple rules that can be used to predict dissociation channels even without calculations.

More detailed calculations for $C_2H_2F_4$ molecule will be a part of our next work.

## 5.  Conclusions and future work.

We performed simulations of the dissociation of 1,1,1,3,3,3-Hexafluoropropane ($C_3H_2F_6$) and 1,3,3,3-Tetrafluoropropene ($C_3H_2F_4$) molecules in their lowest few excited triplet states generated by electron impact.

We run *ab initio* non-adiabatic molecular dynamics with potential energy surfaces and forces calculated "on the fly" by the electronic structure code. The non-adiabatic dynamics calculations are expensive. We demonstrate that, at least for the molecules under consideration, accurate dissociation kinetics can be obtained by running much cheaper molecular dynamics involving only the lowest triplet state. This approach can work even in the case when multiple triplet states are excited initially, e.g., for $C_3H_2F_6$ dissociation, because the molecule ends up in the lowest triplet state within few femtoseconds.

Running the lowest triplet state dynamics on a large number of CPUs, we were able to propagate a large number of trajectories and accumulate good statistics for $C_3H_2F_6$ . We have studied in detail the kinetics of dissociation for C-H, C-C, and C-F bonds at different temperatures, and identified various dissociation channels.

Calculations for $C_3H_2F_4$ demonstrate that the double bonds play an important role, as the triplet excitation is initially localised there.  The excitation then moves towards the neighbouring single bonds, which break more efficiently than the bonds further away from the double bond. The most abundant products detected are produced by breaking of single bonds adjacent to double bonds.  In our next work we will attempt to simulate dissociation patterns of molecules that have been studied experimentally.  The goal of these simulations will be to understand the rules, which should enable us to predict products of dissociation and, therefore, to enable control of chemical composition of plasma. Our effort to find simple rules of dissociation of molecules after electron impact supports that of recent works [26, 27], where dissociation of molecules in low temperature plasma have been investigated experimentally.

*Ab initio* molecular dynamics is computationally expensive, even when only one triplet state is included.  At each time step an electronic structure package is used to calculate potential energy surface and its derivatives which determine forces between the molecules. A much faster approach would be to use analytical force fields similar to those used in molecular dynamics of biomolecules [28]. Force fields for molecules in triplet states obviously differ from those of molecules in their ground electronic state. However, by accumulating more *ab initio* MD data we will try to generate force fields for triplet excited states. Recently an attempt has been made to model triplet state force field with machine learning parametrisation [29], but other parametrisations, similar in spirit to those used for the force fields of ground state [28] may be possible.

Hydrofluorocarbons and other organofluorine molecules are broadly used in the microelectronic industry for various plasma technologies. Recently they came under scrutiny due to their potential environmental damage and the importance of finding new less environmentally damaging molecules has recently been highlighted [26, 27]. We demonstrate that simulations of the dynamics of their dissociation is possible and can be done in the manner similar to how simulations of photodissociation were done previously. Also, perhaps, simple rules which allow predictions of dissociation channels can be found, improving plasma modelling for applications.

Previously, simulations of dynamics involving excited electronic states were mostly focused on the photochemical processes, where singlet states are produced via photon absorption. In this paper we are expanding to the dynamics after excitation by electron impact, which involves low energy triplet states that are energetically separated from higher-lying singlet states. However, the methodology is very much the same for triplet and singlet states. Simulating dissociation of the molecules in triplet states excited by electron impact is actually simpler than that for the photodissociation because potential energy surface of triplet state generally includes repulsive parts. To our best knowledge theoretical simulations of dissociation of molecules in triplet states are very rare. One example is the work [8], where dissociation of methane has been investigated using surface hoping method. Given the practical importance of triplet states dissociation for plasma technologies we plan to continue this work and want to bring this problem to the attention of Molecular Dynamics community.

**Acknowledgements**

We acknowledge EPSRC grant EP/P021123/1 and UKRI grant ST/R005133/1. We also would like to thank Prof Mark Kushner for valuable discussion and encouraging comments.